\documentclass[aps,prb,twocolumn,superscriptaddress,10pt,article,nofootinbib,longbibliography]{revtex4-2}
\usepackage{graphicx,amsfonts,amssymb,amsmath,hyperref,hypcap,enumerate,bm,color,siunitx}

\newcommand{\braket}[2]{\left\langle #1 | #2 \right\rangle}
\newcommand{\bra}[1]{\left\langle#1\right|}
\newcommand{\ket}[1]{\left|#1\right\rangle}

\usepackage{comment}

\newcommand{\comm}[2]{\left[#1,#2\right]}

\newcommand{\half}{\frac{1}{2}}

\newcommand{\beq}{\begin{equation}}
\newcommand{\eneq}{\end{equation}}

\def\mA{{\mathcal{A}}}

\begin{document}

\title{Electrical control of intrinsic nonlinear Hall effect in antiferromagnetic topological insulator sandwiches}

\author{Ruobing Mei}
\affiliation{Department of Physics, The Pennsylvania State University, University Park,  Pennsylvania 16802, USA}
\author{Daniel Kaplan}
\affiliation{Center for Materials Theory, Department of Physics and Astronomy,
Rutgers University, Piscataway, NJ 08854, USA}
\author{Binghai Yan}
\affiliation{Department of Condensed Matter Physics, Weizmann Institute of Science, Rehovot 7610001, Israel}
\author{Cui-Zu Chang}
\affiliation{Department of Physics, The Pennsylvania State University, University Park,  Pennsylvania 16802, USA}
\author{Chao-Xing Liu}
\email{Corresponding author: cxl56@psu.edu}
\affiliation{Department of Physics, The Pennsylvania State University, University Park,  Pennsylvania 16802, USA}

\begin{abstract} 
Nonlinear Hall effect (NHE) can originate from the quantum metric mechanism in antiferromagnetic topological materials with $PT$ symmetry, which has been experimentally observed in MnBi$_2$Te$_4$  \cite{gao2023quantum,wang2023quantum}. In this work, we propose that breaking $PT$ symmetry via external electric fields can lead to a dramatic enhancement of NHE, thus allowing for an electric control of NHE. Microscopically, this is because breaking $PT$ symmetry can lift spin degeneracy of a Kramers' pair, giving rise to additional contributions within one Kramers' pair of bands. We demonstrate this enhancement through a model Hamiltonian that describes an antiferromagnetic topological insulator sandwich structure.
\end{abstract}

\date{\today}


\maketitle



\begin{figure*}
    \centering
    \includegraphics[width=\textwidth]{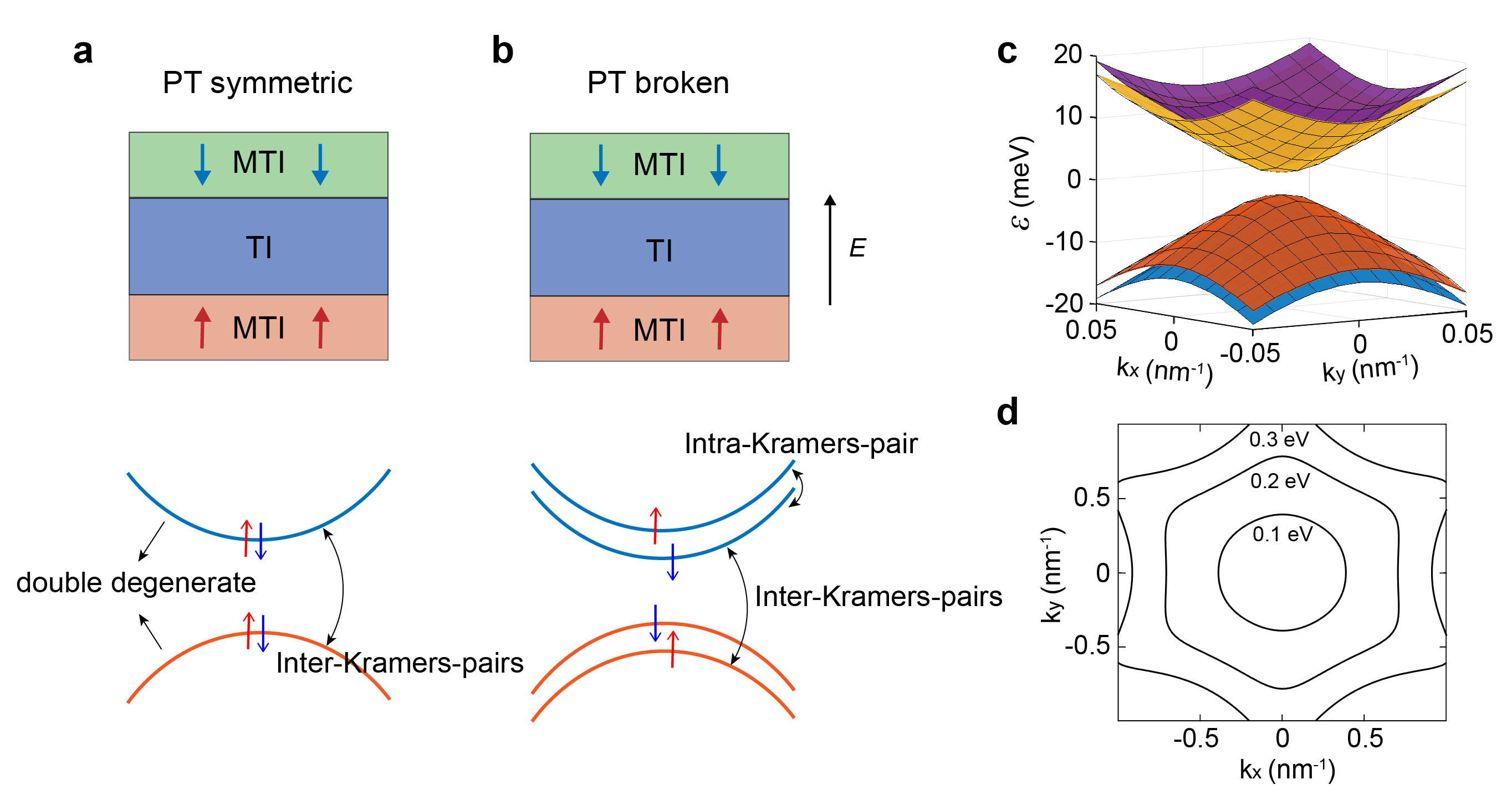}
    \caption{(a) $PT$-symmetric magnetic TI sandwich structure. The magnetizations at the top and bottom surfaces are anti-parallelly aligned. Each band is double-degenerate due to the PT symmetry and the quantum metric dipole only arises between different Kramers' pairs of bands. (b) When an external electric field breaks $PT$ symmetry, the Kramers' pairs split in energy, giving rise to additional contributions between two bands within one Kramers' pair. (c) The energy dispersion for $t = \SI{2}{meV}$ and $V_0 = \SI{1}{meV}$.  (d) Fermi surface contours of the lowest conduction band at the Fermi energies $\varepsilon_f = \SI{0.1}{eV}$, $\SI{0.2}{eV}$, and $\SI{0.3}{eV}$.}
    \label{fig:fig1}
\end{figure*}

{\it Introduction - } Integrating magnetism into topological insulators (TIs) can break time reversal symmetry and lead to the emergence of magnetic topological phases, e.g. the quantum anomalous Hall (QAH) insulators \cite{qi2008topological,yu2010quantized,chang2013experimental,chang2015high,mogi2015magnetic,ou2018enhancing,deng2020quantum,chang2023colloquium} and axion insulators (AIs) \cite{kou2015metal,wang2014universal,xiao2018realization,mogi2017tailoring}. Magnetism has been successfully achieved in TIs by either doping magnetic impurities, e.g. Cr and/or V doped (Bi,Sb)$_2$Te$_3$, or growing stoichiometric antiferromagnetic topological compound, MnBi$_2$Te$_4$. When ferromagnetism is achieved in TI films, both surface states are gapped, leading to the QAH effect \cite{qi2008topological,qi2011topological,chu2011surface,chang2023colloquium}. The quantized Hall response for QAH states have been unambiguously observed in several systems, including magnetically doped TIs \cite{chang2013experimental,chang2015high,mogi2015magnetic,ou2018enhancing}, MnBi$_2$Te$_4$ films \cite{deng2020quantum} and twisted graphene and transition metal dichalcogenides materials \cite{sharpe2019emergent,serlin2020intrinsic,xu2023observation,cai2023signatures,lu2024fractional,park2023observation,tao2024valley,han2024correlated,han2024large,sha2024observation}.
When two surface states of TI films are gapped by antiferromagnetic (AFM) alignment of magnetization, which can be achieved in a magnetic TI sandwich structures with AFM alignment at two surfaces (dubbed as “AFM TI sandwiches” below), as shown in Fig. \ref{fig:fig1}(a), or in even septuple layers (SLs) of MnBi$_2$Te$_4$ films \cite{otrokov2019prediction,gong2019experimental}, the AI phase was theoretically predicted and can host quantized magnetoelectric response \cite{wang2014universal,li2019intrinsic,zhang2019topological,li2022identifying}. Zero Hall plateau observed in these AFM TI sandwiches\cite{xiao2018realization,mogi2017tailoring} provides evidence for the AI phase. Optical experiments have also been studied in AFM TI systems to explore the axion electrodynamics of AI phase \cite{wu2016quantized,dziom2017observation,okada2016terahertz,tse2010giant,sekine2021axion}. 

Recently, quantized topological and non-quantized geometric responses have been generalized to the nonlinear regime in topological materials 
\cite{Holder2020,ma2021topology,de2017quantized,tokura2018nonreciprocal,morimoto2016topological,orenstein2021topology,du2021nonlinear,zuber2021nonlinear,bao2022light,nagaosa2023nonreciprocal}. A notable example is the nonlinear Hall effect (NHE), which describes the Hall current response at the nonlinear order of electric fields. There are two major intrinsic geometric mechanisms for NHE, the Berry curvature dipole \cite{sodemann2015quantum,Zhang2018,ma2019observation,du2018band} and the quantum metric dipole \cite{wang2021intrinsic,kaplan2022unification,gao2014field,liu2021intrinsic}. These two mechanisms have different symmetry properties. The Berry curvature dipole induced NHE can exist in a time reversal ($T$) symmetric system, but is forbidden by $PT$ symmetry, where $P$ is inversion. This is because the $PT$ symmetry guarantees the double degeneracy of all bands at each momentum. Due to its similarity to the Kramers' theorem for the $T$-symmetric systems \cite{ahn2019failure,tasaki2020physics}, these spin degenerate bands are dubbed “Kramers' pairs” below. As a result, the Berry curvature, as well as Berry curvature dipole, has to vanish at each momentum in the Brillouin zone (BZ). In contrast, the quantum metric dipole (also called “intrinsic NHE” \cite{wang2021intrinsic,liu2021intrinsic}) requires the breaking of $T$ symmetry, but it can exist in a $PT$-symmetric system. The AFM TI systems, including AFM TI sandwiches and even SLs of MnBi$_2$Te$_4$ films, can possess $PT$ symmetry but break both $P$ and $T$ symmetry due to the AFM order, and thus provide an ideal platform to examine the intrinsic NHE induced by quantum metric. Recently, intrinsic NHE has been experimentally observed in even SLs of MnBi$_2$Te$_4$ films \cite{gao2023quantum,wang2023quantum}. The quantum metric mechanism requires breaking both $T$ and $P$ but also occurs in $PT$-breaking systems. Our main objective is to understand the dependence of intrinsic NHE on the $PT$ breaking, which can be naturally achieved by an external out-of-plane electric field. 

In this work, we study the NHE in a model Hamiltonian of AFM TI sandwiches under external electric fields in Fig. \ref{fig:fig1}(a) and (b). 
Inversion symmetry breaking by electric fields can lift the spin degeneracy of a Kramers' pair of bands. Although these two spin bands are no longer degenerate in energy, we continue using the terminology ``Kramers' pair'' to denote these two spin bands. In $PT$-symmetric systems, only the contribution between different Kramers' pairs of bands (dubbed “inter-Kramers'-pairs” below) exist for the NHE. Our main result here is to show additional NHE contributions from two spin bands within one Kramers' pair (dubbed “intra-Kramers'-pair” below) can emerge when the inversion symmetry breaking is strong enough so that the energy splitting between two spin bands is larger than disorder broadening. 
Furthermore, we find the intra-Kramers'-pair NHE contribution can dominate over the inter-Kramers'-pairs NHE contribution in the thin film limit when the top and bottom surface states are strongly hybridized. 
Thus, we predict an enhancement of the intrinsic NHE in the magnetic TI sandwiches due to the $PT$ symmetry breaking. 

{\it Model Hamiltonian and symmetry for AFM TI sandwiches -}
We consider a model Hamiltonian for the AFM TI sandwiches, as shown in Fig. \ref{fig:fig1}(a), where the top and bottom surface states of the TI layers open the gaps with opposite signs. We assume the Fermi energy of this sandwich structure is within the TI bulk, so the low-energy effective Hamiltonian for two surface states reads
\beq \label{eq:Ham}
\begin{aligned}
 H =& v_f(k_y \sigma_x - k_x \sigma_y) \tau_z + m\sigma_z \tau_z  \\
 &+ \lambda (k_+^3 + k_-^3) \sigma_z \tau_z + t\tau_x + V_0 \tau_z 
\end{aligned}
\eneq
where $\sigma_i$ and $\tau_i$ ($i=x,y,z$) are Pauli matrices in the spin and surface states basis, $v_f$ is the Fermi velocity, $m$ is the exchange coupling strength, $\lambda$ is the hexagonal warping coefficient \cite{fu2009hexagonal}, and $k_{\pm} = k_x\pm ik_y$. We choose the parameters to be $v_f = \SI{2.55}{eV\angstrom}$, $\lambda = \SI{125}{eV\angstrom^3}$, and $m = \SI{1}{meV}$ \cite{fu2009hexagonal}. $V_0$ is the asymmetric potential created by the out-of-plane electric field $E$, $V_0=eEd$, where $d$ is the TI film thickness. The coupling parameter $t$ describes the hybridization strength between the top and bottom surface states, which quickly decays as the TI film thickness $d$ increases and is chosen in the range of $0 \sim 10$ meV\cite{zhang2010crossover,sakamoto2010spectroscopic,wang2019dimensional,fang2020layer}.

The model Hamiltonian breaks $P = \tau_x$ and $T = i\sigma_y K$ symmetries, and preserves $C_{3z} = e^{-i\frac{\pi}{3}\sigma_z}$ and $M_xT = -i\sigma_zK$ symmetries. In particular, the exchange coupling term breaks $P$, $T$, $M_x=i\sigma_x$, and $M_y = i\sigma_y$ symmetries, the hexagonal warping term breaks the full rotational symmetries down to the $C_{3z}$ symmetry, and the asymmetric potential term $V_0$ breaks $P$ and $PT$ symmetries. When the asymmetric potential is absent, the system is $PT$ symmetric. The symmetry properties of this system also give a strong constraint on the form of nonlinear conductivity $\sigma_{abc}$, defined by $j^a = \sigma_{abc} E^b E^c$, where $a,b,c=x,y$. 

By performing the symmetry analysis (see Appendix B \cite{sm2024}), we find that $M_x T$ requires $\sigma_{yyy} = \sigma_{yxx} = \sigma_{xyx} = \sigma_{xxy} = 0$ while $C_{3z}$ requires $\sigma_{xyy} = \sigma_{yxy} =\sigma_{yyx} = -\sigma_{xxx}$. Therefore, $\sigma_{xyy}$ is the only independent component for this model.

\begin{figure*}
    \centering
    \includegraphics[width=\textwidth]{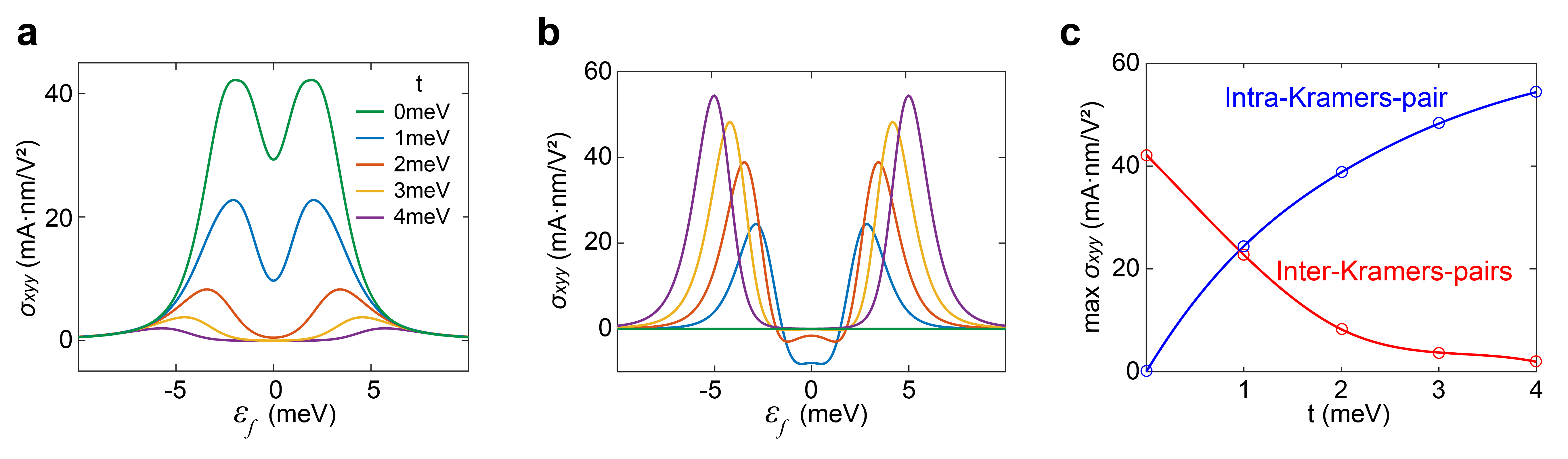}
    \caption{(a) $\sigma^{\text{inter}}_{xyy}$ as a function of Fermi energy $\varepsilon_f$ for the inter-Kramers'-pairs contribution at $V_0 = \SI{2}{meV}$, $\Gamma = \SI{0.6}{meV}$, and $t = \SI{0}{meV}$, $\SI{1}{meV}$, $\SI{2}{meV}$, $\SI{3}{meV}$ and $\SI{4}{meV}$. (b) $\sigma^{\text{intra}}_{xyy}$ as a function of Fermi energy $\varepsilon_f$ for the intra-Kramers'-pair contribution with different $t$. (c) The maximum value of $\sigma^{\text{inter}}_{xyy}$ and $\sigma^{\text{intra}}_{xyy}$ respect to Fermi energy $\varepsilon_f$ as a function of coupling coefficient $t$.}
    \label{fig:fig2}
\end{figure*}

The eigen-energies of the Hamiltonian can be solved as $\varepsilon_{n\mu} = n \sqrt{t^2 + (A + \mu V_0)^2}$, where $n, \mu = \pm$ and $A = \sqrt{v_f^2 k^2 + (m + 2 \lambda k_x (k_x^2-3k_y^2))^2}$, where $n$ is the index for different sets of Kramers' pairs and $\mu$ labels two spin states in one Kramer pair. When $V_0 = 0$, there are two sets of degenerate Kramers' pairs with the eigen-energies $\varepsilon_{\pm} = \pm \sqrt{A^2 +t^2}$. Such degeneracy is broken by a nonzero $V_0$, as shown in Fig. \ref{fig:fig1}(c), where the energy dispersion is calculated for $t = \SI{2}{meV}$ and $V_0 = \SI{1}{meV}$. Fig. \ref{fig:fig1}(d) depicts the Fermi surface contours of the lowest conduction band at different energies, where the hexagonal warping effect is visible for a large momentum $k$.

{\it Quantum-metric-induced NHE in $PT$-symmetric and $PT$-breaking systems - }
Since the Berry curvature dipole is forbidden by $C_{3z}$ \cite{sodemann2015quantum,suzuki2017cluster}, it is excluded in our model even when $PT$ symmetry is broken. Therefore, we focus on the intrinsic NHE. As described in Appendix A\cite{sm2024}, the intrinsic nonlinear Hall conductivity the can be written as
\beq \label{eq:crossover}
\begin{aligned}
\sigma_{xyy} = &-\frac{e^3}{\hbar} \sum_{n\mu,m\nu} \int \frac{d^3k}{(2\pi)^3} f_{n \mu} \Big[ \partial_x (\alpha_{n\mu,m\nu} g^{yy}_{n\mu,m\nu}) \\
&+  \partial_y (\beta_{n\mu,m\nu} g^{xy}_{n\mu,m\nu}) \Big],
\end{aligned}
\eneq
where $\partial_a = \frac{\partial}{\partial k_a}$ and $f_{n\mu} = \frac{1}{e^{(\varepsilon_{n\mu}-\varepsilon_f)/\Gamma}+1}$ is the Fermi distribution function, with the eigen-energy $\varepsilon_{n\mu}$, the Fermi energy $\varepsilon_f$, and the temperature broadening $\Gamma = k_B T$. The band-resolved quantum metric reads \cite{kaplan2022unification}
\beq
g^{ab}_{n\mu,m\nu} = \mA^a_{n\mu,m\nu} \mA^b_{m\nu,n\mu}+\mA^b_{n\mu,m\nu} \mA^a_{m\nu,n\mu},
\eneq
where $\mA^a_{n\mu,m\nu} = \braket{n\mu}{i\partial_a|m\nu}$ is the Berry connection, and $\ket{n\mu}$ is the eigen-wavefunction of the Hamiltonian. The remaining dependence on disorder broadening $\Delta_\tau=\hbar/\tau$ with the relaxation time $\tau$ in the intrinsic conductivity is
via the functions $\alpha$ and $\beta$, defined as  
\beq \label{eq:alpha}
\alpha_{n\mu,m\nu} = \text{Re} \Big[ \frac{\varepsilon_{n\mu,m\nu}}{\Delta_{\tau} (i \varepsilon_{n\mu,m\nu}+\Delta_{\tau})} - (n\mu \leftrightarrow m\nu) \Big],
\eneq
\beq \label{eq:beta}
\begin{aligned}
\beta_{n\mu,m\nu} = \text{Re} &\Big[\frac{\varepsilon_{n\mu,m\nu}}{i \varepsilon_{n\mu,m\nu} + \Delta_{\tau}/2} \left(\frac{1}{i \varepsilon_{n\mu,m\nu} + \Delta_{\tau}} + \frac{1}{\Delta_{\tau}}\right) \\ 
&- (n\mu \leftrightarrow m\nu) \Big]. 
\end{aligned}
\eneq
This dependence becomes significant whenever the energy difference $\varepsilon_{n\mu,m\nu} = \varepsilon_{n\mu} - \varepsilon_{m\nu}$ is of the order of $\Delta_\tau$.

For the $PT$-symmetric case, the two spin bands within one Kramers' pair ($n = m$) are degenerate so that $\varepsilon_{n\mu,n\nu} = 0$ and thus $\alpha_{n\mu,n\nu} = \beta_{n\mu,n\nu} = 0$. Therefore, the degenerate states within one Kramers' pair give no contribution. We consider the case when the energy difference between different sets of Kramers' pairs is much larger that the disorder level, i.e. $\varepsilon_{n\mu,m\nu} \gg \Delta_{\tau}$ ($n \neq m$), and can apply the expansion 
\beq \label{eq:approximation}
\frac{1}{i \varepsilon_{n\mu,m\nu} + \Delta_{\tau}} = \frac{1}{i \varepsilon_{n\mu,m\nu}} + 
\frac{\Delta_{\tau}}{\varepsilon^2_{n\mu,m\nu}} + \mathcal{O} (\Delta^2_{\tau}). 
\eneq
Up to the first order in $\Delta_{\tau}$, we find $\alpha_{n\mu,m\nu} = \frac{2}{\varepsilon_{n\mu,m\nu}}$ and $\beta_{n\mu,m\nu} = -\frac{1}{\varepsilon_{n\mu,m\nu}}$ for $n\neq m$, which leads to
\beq \label{eq:Hall_conductivity}
\sigma_{xyy} = - \frac{e^3}{\hbar} \sum_{n\mu} \int_k f_{n\mu} (2\partial_x G^{yy}_{n\mu} - \partial_y G^{xy}_{n\mu}),
\eneq
where $G^{ab}_{n\mu} = \sum_{m\nu} g^{ab}_{n\mu,m\nu}/\varepsilon_{n\mu,m\nu}$ with $m\neq n$. 
The derived NHE expression is only for the quantum metric between non-degenerate bands (“inter-Kramers'-pairs”), and is consistent with Ref. \cite{kaplan2022unification}. The result can be also be viewed from the point of view of semiclassical theory. The origin of the NHE is traced to systematic corrections to the Berry curvature and energy dispersion stemming  from the dressing of operators due to the electric field. Semiclassically, $H \to H + e \mathbf{E} \cdot \mathbf{r}$. Invoking the $U(1)^{N}$ symmetry of the Block Hamiltonian, it is possible to remove the gauge dependent linear coupling via a unitary transformation $H' =  e^{-S}H e^{S}$, with $S$ fixed to remove the linear in $\mathbf{E}$ contribution (see Ref. \cite{kaplan2022unification}). However, it should be noted that the semiclassical approach fixes the diagonal components of $H'$. It is also possible to construct off-diagonal elements of $H'$ and hence $v' = \frac{\partial H'}{\partial \mathbf{k}}$. By allowing off-diagonal elements in the Boltzmann equation away from the clean limit, one recovers Eqs. \eqref{eq:alpha} and \eqref{eq:beta}. Importantly, in the limit where the relaxation $\tau$ only enters the diagonal part of the density fucntion $f_{n\mu}$(the standard semiclassical assumption), the result of the Boltzmann treatment and the approach presented here coincide completely.

In $PT$-breaking systems, the asymmetric potential $V_0$ splits the energy of two spin bands within one Kramers' pair. 
When the energy difference between these two spin bands is much larger than the disorder level, i.e. $V_0 \gg \Delta_{\tau}$, the relaxation time approximation [Eq. \eqref{eq:approximation}] is valid for any pairs of $(n,\mu)$ and $(m,\nu)$ and thus $\alpha_{n\mu,m\nu} = \frac{2}{\varepsilon_{n\mu,m\nu}}$ and $\beta_{n\mu,m\nu} = -\frac{1}{\varepsilon_{n\mu,m\nu}}$ for both intra-Kramers' pairs ($n=m$) and inter-Kramers' pairs ($n\neq m$). We then 
obtain a similar expression as Eq. \eqref{eq:Hall_conductivity}, but the summation over $m$ in $G^{ab}_{n\mu} = \sum_{m\nu} g^{ab}_{n\mu,m\nu}/\varepsilon_{n\mu,m\nu}$ should also includes $m=n$. Therefore, the quantum metric within one Kramers' pair of bands (“intra-Kramers'-pair”) can also contribute to the NHE in addition to the inter-Kramers'-pairs part. 

\begin{figure*}
    \centering
    \includegraphics[width=\textwidth]{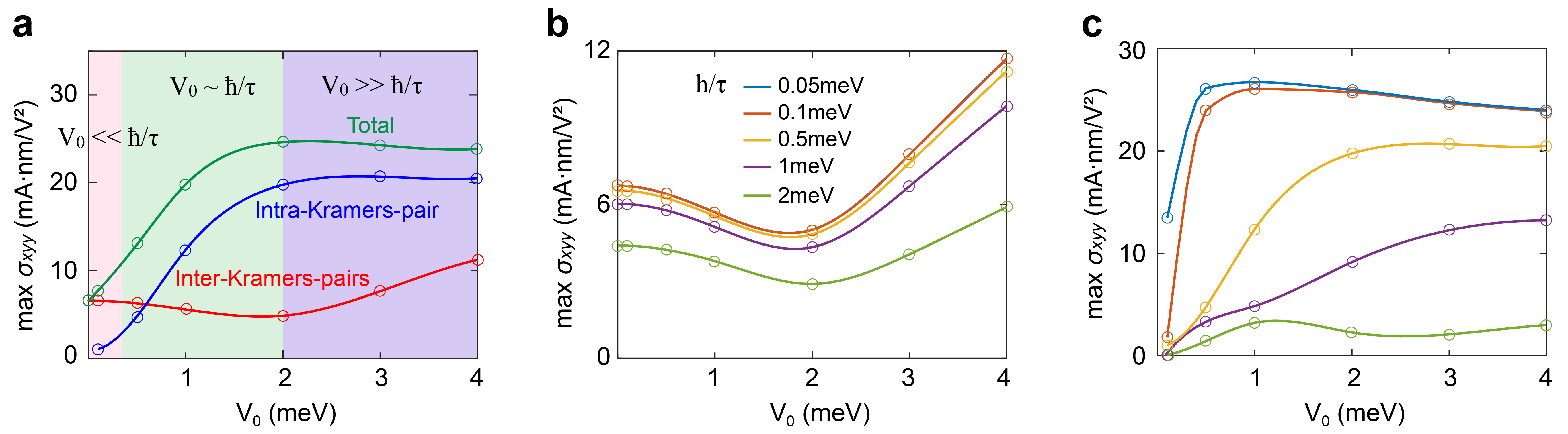}
    \caption{(a) The maximum value of $\sigma_{xyy}$ for the intra-Kramers'-pair (blue), inter-Kramers'-pairs (red), and total (green) contributions as a function of asymmetric potential $V_0$ at $t=\SI{2}{meV}$ and $\hbar/\tau = \SI{0.5}{meV}$.  (b) The maximum value of $\sigma^{\text{inter}}_{xyy}$ as a function of asymmetric potential $V_0$ at different disorder levels $\hbar/\tau$ for the inter-Kramers'-pairs contribution at $t=\SI{2}{meV}$. (c) The maximum value of $\sigma^{\text{intra}}_{xyy}$ as a function of $V_0$ for the intra-Kramers'-pair contribution at $t = \SI{2}{meV}$.}
    \label{fig:fig3}
\end{figure*}

{\it Electric field control of NHE - } 
We numerically evaluate the NHE for the model Hamiltonian [Eq. \eqref{eq:Ham}] based on Eq. \eqref{eq:crossover}. Fig. \ref{fig:fig2}(a) and (b) shows the Fermi energy $\varepsilon_f$ dependence of the inter-Kramers'-pair contribution $\sigma^{\text{inter}}_{xyy}$ and the intra-Kramers'-pair contribution $\sigma^{\text{intra}}_{xyy}$ at different coupling coefficient $t$, respectively, at $V_0 = \SI{2}{meV}$, $\Gamma = \SI{0.6}{meV}$, and assuming the disorder level is very low ($V_0 \gg \hbar/\tau$). For both components, $\sigma_{xyy}$ displays the same sign in the electron and hole doping regions. The intra-Kramers'-pair contribution $\sigma^{\text{intra}}_{xyy}$ vanishes for $t=\SI{0}{meV}$, which indicates that the inter-layer hybridization term is crucial for non-zero quantum metric between two bands in one Kramers' pair. Furthermore, $\sigma^{\text{intra}}_{xyy}$ increases with $t$ while the inter-Kramers'-pairs contribution $\sigma^{\text{inter}}_{xyy}$ decreases, as illustrated in Fig. \ref{fig:fig2}(c), which plots the maximum values max($\sigma^{\text{intra}}_{xyy}$) and max($\sigma^{\text{inter}}_{xyy}$) as a function of the coupling coefficient $t$. Here the maximum refers to the peak value of $\sigma^{\text{intra}}_{xyy}$ and $\sigma^{\text{inter}}_{xyy}$ when varying with the Fermi energy $\varepsilon_f$ in Fig. \ref{fig:fig2}(a) and (b). Therefore, the intra-Kramers'-pair contribution $\sigma^{\text{intra}}_{xyy}$ plays a more important role when the inter-layer hybridization is stronger, i.e. when the sample thickness is thinner. 

We summarize max($\sigma_{xyy}$) as a function of $V_0$ in Fig. \ref{fig:fig3}(a), in which the green curve depicts the variation of total max($\sigma_{xyy}$), while the blue and red curves illustrate max($\sigma^{\text{intra}}_{xyy}$) and max($\sigma^{\text{inter}}_{xyy}$), respectively, at $t=\SI{2}{meV}$ and $\hbar/\tau = \SI{0.5}{meV}$. The intra-Kramers'-pair contribution max($\sigma^{\text{intra}}_{xyy}$) is zero at $V_0 \sim 0$, and increases rapidly when $V_0$ is increasing. In contrast, the inter-Kramers'-pairs contribution max($\sigma^{\text{inter}}_{xyy}$) almost remains a constant with increasing $V_0$. Thus, one can divide the variation of max($\sigma_{xyy}$) into three regions for a fixed disorder strength. When $V_0 \ll \hbar/\tau$, max($\sigma^{\text{intra}}_{xyy}$) is close to zero and max($\sigma^{\text{inter}}_{xyy}$) dominates. When $V_0 \sim \hbar/\tau$, max($\sigma^{\text{intra}}_{xyy}$) increases rapidly with $V_0$, while max($\sigma^{\text{inter}}_{xyy}$) exhibits a small decrease. When $V_0 \gg \hbar/\tau$, max($\sigma^{\text{intra}}_{xyy}$) saturates while max($\sigma^{\text{inter}}_{xyy}$) reveals an upturn. 
Fig. \ref{fig:fig3}(b) and (c) show the max($\sigma^{\text{inter}}_{xyy}$) and max($\sigma^{\text{intra}}_{xyy}$) as a function of $V_0$ for different $\hbar/\tau$, from which one can see the above scenario for the division of three regions generally remains valid for different $\hbar/\tau$ values. Furthermore, with increasing $\hbar/\tau$, we find both max($\sigma_{xyy}^{\text{inter}}$) and max($\sigma_{xyy}^{\text{intra}}$) are reduced.

{\it Conclusion -}
To summarize, we showed the enhancement of the intrinsic NHE in AFM TI sandwiches via the breaking of inversion symmetry. The intrinsic NHE has been observed for even SLs of MnBi$_2$Te$_4$ in two recent experiments \cite{gao2023quantum,wang2023quantum}, and the displacement field dependence of the NHE has been measured. As discussed previously, the amount of enhancement is affected by the film thickness as well as the disorder level in the system. 
In Ref. \cite{gao2023quantum}, a relative small enhancement of the NHE was reported as the displacement field increases. The conductivity in this experiment is $\sigma_{xx} \approx \SI{14}{mS}$ at carrier density $n_e = 3\times 10^{12} \text{cm}^{-2}$, and from $\tau = \frac{m^* \sigma_{xx}}{e^2 n_e}$ with the electron effective mass $m^* \approx 0.1 m_e$ \cite{an2021nanodevices}, we estimate the disorder level to be $\hbar/\tau \sim \SI{0.4}{meV}$. However, the experiment was perform in 6 SL MnBi$_2$Te$_4$, of which the hybridization strength between two surface states is considerably small \cite{zhang2010crossover,sakamoto2010spectroscopic,wang2019dimensional,fang2020layer}, thus suppressing the intra-Kramers'-pair contribution, as well as the enhancement of NHE. 
In Ref.~\cite{wang2023quantum}, on the other hand, very little or no increase of the NHE was observed in the displacement field dependence measurement. We estimate the disorder level as $\hbar/\tau \sim \SI{20}{meV}$ with $\sigma_{xx}/n_e \approx 9 \times 10^{-11} \mu \text{S}\, \text{cm}^2$ and strong disorder scattering can greatly suppress the NHE enhancement. 

The scaling analysis between the nonlinear Hall conductivity and the longitudinal conductivity by varying temperatures has been used to distinguish the intrinsic NHE, which is the sole contribution that is independent of relaxation time $\tau$ in the weak disorder limit, from other extrinsic mechanism, e.g. skew scattering and side jump\cite{wang2021intrinsic,wang2023quantum,du2019disorder,nandy2019symmetry,du2021quantum}, with strong dependence on $\tau$. As our NHE formula Eq. \eqref{eq:crossover} is beyond the weak disorder limit, the relaxation time $\tau$ dependence of the intrinsic NHE is found in Fig. \ref{fig:fig3}(b) and (c), when the disorder broadening is comparable to band energy splitting. We note in both experiments \cite{gao2023quantum,wang2023quantum} the NHE shows very little dependence of $\tau$ when varying temperatures. As the NHE only appears below the Neel temperature $\sim \SI{24}{K}$ \cite{he2020mnbi2te4}, we can estimate the change of relaxation time to be $\frac{\delta \tau}{\tau}\sim 20\%$ within this temperature range. From \ref{fig:fig3}(b) and (c), a significant change of NHE can only occur when $\tau$ is changed by several times. Thus, $20\%$ variation of $\tau$ can only lead a negligible change of the NHE, while a more feasible control of the NHE is through a displacement field, which was indeed observed experimentally \cite{gao2023quantum}. 

In our calculations, the value of NHE is around the order of $10\, \text{mA} \, \text{nm}/\text{V}^2$, while the experimental values reported in Refs. \cite{gao2023quantum} and \cite{wang2023quantum} are both $\sim 100\, \text{mA} \, \text{nm}/\text{V}^2$, one order larger than the calculated value. To explain this discrepancy, it was proposed that the quantum metric in experiments might be enhanced by the modified surface band structure of MnBi$_2$Te$_4$ \cite{wang2023quantum}. As disorder scattering is strong in Ref. \cite{wang2023quantum}, a full quantum mechanical treatment of disorder effect beyond the relaxation time approximation \cite{xiao2019theory,du2019disorder,nandy2019symmetry,ortix2021nonlinear} is required. Furthermore, the edge transport may also give rise to the NHE for the Fermi energy close to the band edges \cite{yasuda2020large,zhang2022controlled}, which is beyond the current theoretical formalism. 

{\it Acknowledgement -}
We thank Weibo Gao for helpful discussion. R.B.M., C.Z.C. and C.-X.L. acknowledge the support from the NSF through The Pennsylvania State University Materials Research Science and Engineering Center [DMR-2011839]. C.Z.C. and C.-X.L. also acknowledge support from NSF grant via the grant number DMR-2241327. C. -Z. C. acknowledges the support from the Gordon and Betty Moore Foundation’s EPiQS Initiative (Grant GBMF9063 to C. -Z. C.). D. K. is supported by the Abrahams Postdoctoral Fellowship of the Center for Materials Theory, Rutgers University and the Zuckerman STEM Fellowship. B.Y. acknowledges the financial support by the European Research Council (ERC Consolidator Grant ``NonlinearTopo'', No. 815869) and the Israel Science Foundation (ISF: 2932/21, 2974/23).


\appendix

\renewcommand\thefigure{\thesection.\arabic{figure}} 
\setcounter{figure}{0}

\section{Formalism of nonlinear Hall effect}

We follow the density matrix formalism \cite{watanabe2020nonlinear} for the derivation of the nonlinear Hall conductivity. We start from the density matrix $\rho = \rho^{(0)} + \rho^{(1)} + \rho^{(2)} + \cdots$ and the Liouville equation
\beq
  \partial_t \rho= - \frac{i}{\hbar} \comm{H}{\rho}-\frac{\rho-\rho^{(0)}}{\tau}.
\eneq
We consider the Hamiltonian $H=H_0+H_1$ with $H_1=e\Vec{E}\cdot\Vec{\mathcal{R}}$, where the position operator $\Vec{\mathcal{R}}$ acts on an operator $\mathcal{O}$ as
$i\comm{\Vec{\mathcal{R}}}{\mathcal{O}}_{nm}=\nabla_{\Vec{k}}\mathcal{O}_{nm}-i\comm{\Vec{\mA}}{\mathcal{O}}_{nm}$ and $\Vec{\mA}_{nm}=i\bra{n} \Vec{\partial} \ket{m}$ is the Berry connection. 
At the zeroth order, $\dot{\rho}^{(0)}=-\frac{i}{\hbar} \comm{H_0}{\rho^{(0)}}$ and $\rho^{(0)}=\sum_{n\mu} f_{n\mu} \ket{n\mu}\bra{n\mu}$, where $\ket{n\mu}$ are the eigenvectors of $H_0$ and $f_{n\mu}$ is the Fermi distribution function. Here $n$ is the index for different sets of Kramers' pairs and $\mu$ is the index for the two spin bands within one Kramers' pair.

At the first order, we have
\beq
  \partial_t\rho^{(1)}=-\frac{i}{\hbar} \comm{H_1}{\rho^{(0)}}-\frac{i}{\hbar} \comm{H_0}{\rho^{(1)}}-\frac{\rho^{(1)}}{\tau_1},
\eneq

\beq
\hspace*{-8mm} 
\frac{i}{\hbar} \comm{H_0}{\rho^{(1)}}_{n\mu,m\nu}+\frac{\rho^{(1)}_{n\mu,m\nu}}{\tau_1}=-\frac{i}{\hbar}\comm{H_1}{\rho^{(0)}}_{n\mu,m\nu}
 \hspace{-30pt}
\eneq

\beq \label{eq:rho1}
\begin{aligned}
 \rho^{(1)}_{n\mu,m\nu}= &-\frac{e \tau_1}{\hbar} E^a \partial_a f_{n\mu} \delta_{nm} \delta_{\mu\nu} \\
 & + \frac{ieE^a(f_{m\nu,n\mu}) \mA^a_{n\mu,m\nu}}{i\varepsilon_{n\mu,m\nu}+\hbar/\tau_1},
\end{aligned}
\eneq
where $f_{n\mu,m\nu} = f_{n\mu} - f_{m\nu}$, $\varepsilon_{n\mu,m\nu} = \varepsilon_{n\mu}-\varepsilon_{m\nu}$, $\mA^a_{n\mu,m\nu} = i \bra{n\mu} \partial_a \ket{m\nu}$, and $\varepsilon_{n\mu}$ is the eigenvalue of $H_0$.

At the second order, we get
\beq
  \partial_t\rho^{(2)}=-\frac{i}{\hbar} [H_1,\rho^{(1)}]-\frac{i}{\hbar} [H_0,\rho^{(2)}]-\frac{\rho^{(2)}}{\tau_2},
\eneq
\beq
\hspace*{-8mm} 
\frac{i}{\hbar} \comm{H_0}{\rho^{(2)}}_{n\mu,m\nu}+\frac{\rho^{(2)}_{n\mu,m\nu}}{\tau_2}=-\frac{i}{\hbar}\comm{H_1}{\rho^{(1)}}_{n\mu,m\nu}
 \hspace{-30pt}
\eneq
\beq \label{eq:density_matrix_sec}
  \rho^{(2)}_{n\mu,m\nu} = \frac{-eE^b\partial_b\rho^{(1)}_{n\mu,m\nu}+ieE^b[\mA^b,\rho^{(1)}]_{n\mu,m\nu}}{i\varepsilon_{n\mu,m\nu}+\hbar/\tau_2}.
\eneq
It should be noted that we use $\tau_1$ and $\tau_2$ to label the relaxation time at the first and second orders, which can generally be two independent parameters. 

By using Eq. \eqref{eq:rho1} and only keeping the contribution at the Fermi surface, we have
\beq \label{eq:ferimi-surface}
\begin{aligned}
&\rho^{(2)}_{n\mu,m\nu} 
= e^2 E^a E^b \Big[ \frac{\tau_1 \tau_2}{\hbar^2} \partial_a \partial_b f_{n\mu} \delta_{nm} \delta_{\mu \nu} \\
& \qquad \quad - \frac{i \tau_1 (\partial_a f_{m\nu,n\mu}) \mA^b_{n\mu,m\nu}}{\hbar (i \varepsilon_{n\mu,m\nu}+\hbar/\tau_2)} \\
& \qquad \quad - \frac{i (\partial_b f_{m\nu,n\mu}) \mA^a_{n\mu,m\nu}}{(i \varepsilon_{n\mu,m\nu}+\hbar/\tau_1) (i \varepsilon_{n\mu,m\nu}+\hbar/\tau_2)} \\
&- \frac{\tau_2}{\hbar}\delta_{nm} \delta_{\mu\nu} \sum_{l\xi} \Bigl( \frac{f_{n\mu,l\xi} \mA^a_{l\xi,n\mu} \mA^b_{n\mu,l\xi}}{i \varepsilon_{n\mu,l\xi}+\hbar/\tau_1} - (n\mu \leftrightarrow l\xi)\Bigr) \Big]. \hspace{-30pt}
\end{aligned}
\eneq
We then derive the current as 
\beq
j^a = -e\sum_{nm,\mu\nu} \int_k \rho^{(2)}_{n\mu,m\nu} v^a_{m\nu,n\mu}, 
\eneq
where $v^a_{n\mu,m\nu} = \bra{n\mu} \partial_a H_0 \ket{m\nu}$ is the velocity operator. The second-order conductivity is defined as $\sigma_{abc} = \frac{j^a}{E^b E^c}$. After symmetrizing $b \leftrightarrow c$ and choosing $\tau = \tau_1 = \tau_2/2$ \cite{kaplan2022unification,kaplan2023unifying,passos2018nonlinear}, we obtain
\beq \label{eq:A10}
\begin{aligned}
\sigma_{abc} &= - \frac{e^3 \tau^2}{\hbar^3} \sum_{n\mu} \int_k f_{n\mu} \partial_a \partial_b \partial_c \varepsilon_{n\mu} \\
&- \frac{e^3}{2\hbar} \sum_{n\mu,m\nu} \int_k [(\partial_b f_{n\mu}) \gamma_{n\mu,m\nu} \Omega^{ca}_{n\mu,m\nu} + (b \leftrightarrow c) ]\hspace{-30pt}  \\ 
&+\frac{e^3}{\hbar} \sum_{n\mu,m\nu} \int_k  \Big[ (\partial_a f_{n\mu}) \alpha_{n\mu,m\nu} g^{bc}_{n\mu,m\nu}  \\
&+ \half \left( (\partial_b f_{n\mu}) \beta_{n\mu,m\nu} g^{ac}_{n\mu,m\nu} + (b \leftrightarrow c) \right) \Big],
\end{aligned}
\eneq
where
\beq
\Omega^{ab}_{n\mu,m\nu} = i (\mA^a_{n\mu,m\nu} \mA^b_{m\nu,n\mu} - \mA^b_{n\mu,m\nu} \mA^a_{m\nu,n\mu}),
\eneq
\beq
g^{ab}_{n\mu,m\nu} = \mA^a_{n\mu,m\nu} \mA^b_{m\nu,n\mu} + \mA^b_{n\mu,m\nu} \mA^a_{m\nu,n\mu}
\eneq
are the band-resolved Berry curvature and quantum metric. The other functions $\alpha_{n\mu,m\nu}$ and $\beta_{n\mu,m\nu}$ are defined in Eqs. \eqref{eq:alpha} and \eqref{eq:beta}, and
\beq
\begin{aligned}
\gamma_{n\mu,m\nu} = \text{Im} &\Big[\frac{\varepsilon_{n\mu,m\nu}}{i \varepsilon_{n\mu,m\nu} + \Delta_{\tau}/2} \left(\frac{1}{i \varepsilon_{n\mu,m\nu} + \Delta_{\tau}} + \frac{1}{\Delta_{\tau}}\right) \\ 
&+ (n\mu \leftrightarrow m\nu) \Big],
\end{aligned}
\eneq
with $\Delta_{\tau} = \hbar/\tau$. In Eq. \eqref{eq:A10}, the first term is called the nonlinear Drude weight; the second term is the Berry curvature dipole induced NHE; the third term is the quantum metric dipole induced NHE, which corresponds to the Eq. \eqref{eq:Hall_conductivity} in the main text.

In the clean limit, i.e. $\tau \rightarrow \infty$, we use the expansion
\beq \label{eq:expansion2}
\frac{1}{i \varepsilon_{n\mu,m\nu} +\hbar/\tau} = \frac{1}{i \varepsilon_{n\mu,m\nu}} + \frac{\tau}{\hbar \varepsilon^2_{n\mu,m\nu}} + \mathcal{O} (\tau^{-2}),
\eneq
For the PT-symmetric case, there is 
\beq \label{eq:alpha_beta_gamma}
\begin{aligned}
\alpha_{n\mu,m\nu} &= \frac{2}{\varepsilon_{n\mu,m\nu}}, \\
\beta_{n\mu,m\nu} &= -\frac{1}{\varepsilon_{n\mu,m\nu}}, \\
\gamma_{n\mu,m\nu} &= - \frac{2\tau}{\hbar},
\end{aligned}
\eneq
for $m\neq n$ and $\alpha_{n\mu,n\nu} =\beta_{n\mu,n\nu} =\gamma_{n\mu,n\nu} = 0$ for $m=n$, while in the $PT$-breaking case, Eq. \eqref{eq:alpha_beta_gamma} is true for both $m\neq n$ and $m=n$. By performing integrating by parts, the nonlinear conductivity reads
\beq
\begin{aligned}
&\sigma_{abc} = - \frac{e^3 \tau^2}{\hbar^3} \sum_{n\mu} \int_k f_{n\mu} \partial_a \partial_b \partial_c \varepsilon_{n\mu} \\
&- \frac{e^3 \tau}{\hbar^2} \sum_{n\mu} \int_k f_{n\mu} (\partial_b \Omega^{ca}_{n\mu} + \partial_c \Omega^{ba}_{n\mu}) \\
&- \frac{e^3}{\hbar} \sum_{n\mu} \int_k f_{n\mu} (2\partial_a G^{bc}_{n\mu} -\half (\partial_b G^{ac}_{n\mu} + \partial_c G^{ab}_{n\mu})),
\end{aligned}
\eneq
where $\Omega^{ab}_{n\mu} = \sum_{m\nu} \Omega^{ab}_{n\mu,m\nu}$ and the normalized quantum metric $G^{ab}_{n\mu} = \sum_{m\nu} g^{ab}_{n\mu,m\nu}/\varepsilon_{n\mu,m\nu}$ with $m\neq n$ in the $PT$-symmetric system. We can see that the nonlinear Drude term is quadratic in the relaxation time $\tau$, the Berry curvature dipole term is linear in $\tau$, and the quantum metric dipole term is independent of $\tau$.

\section{Symmetry analysis of nonlinear conductivity}
The second-order conductivity tensor is defined as
\beq
\begin{aligned}
T_x = \sigma_{xbc} =
\begin{pmatrix}
\sigma_{xxx} & \sigma_{xxy} \\
\sigma_{xyx} & \sigma_{xyy} 
\end{pmatrix}, \\
T_y = \sigma_{ybc} =
\begin{pmatrix}
\sigma_{yxx} & \sigma_{yxy} \\
\sigma_{yyx} & \sigma_{yyy} 
\end{pmatrix}.
\end{aligned}
\eneq
Under an orthogonal transformation matrix $R_{ij}$, the conductivity tensor transforms as $T_i \rightarrow R_{ij} (R T_{j} R^T)$ \cite{utermohlen2021symmetry}. Thus, $C_{3z}$ symmetry whose transformation matrix is
\beq
C_{3z} = 
\begin{pmatrix}
\cos\frac{2\pi}{3} & \sin\frac{2\pi}{3} \\
-\sin\frac{2\pi}{3} & \cos\frac{2\pi}{3}
\end{pmatrix},
\eneq
imposes the constraints $T_i = (C_{3z})_{ij} (C_{3z} T_j C_{3z}^T)$, or explicitly, $\sigma_{xxx}= -\sigma_{xyy} = -\sigma_{yxy} = -\sigma_{yyx}$ and $\sigma_{yyy} = -\sigma_{yxx} = -
\sigma_{xyx} = -\sigma_{xxy}$.

Under $M_xT$ symmetry, the electric field transforms as $E_x \rightarrow -E_x$ and $E_y \rightarrow E_y$, and the current transforms as $j_x \rightarrow j_x$ and $j_y \rightarrow -j_y$. Thus, $M_xT$ symmetry imposes the constraints $\sigma_{yyy} = \sigma_{yxx} = \sigma_{xyx} =\sigma_{xxy} = 0$. Therefore, $\sigma_{xyy}$ is the only independent component of nonlinear conductivity.

\section{Fermi sphere contribution to the nonlinear Hall conductivity}

\begin{figure}
    \centering
    \includegraphics[width=\linewidth]{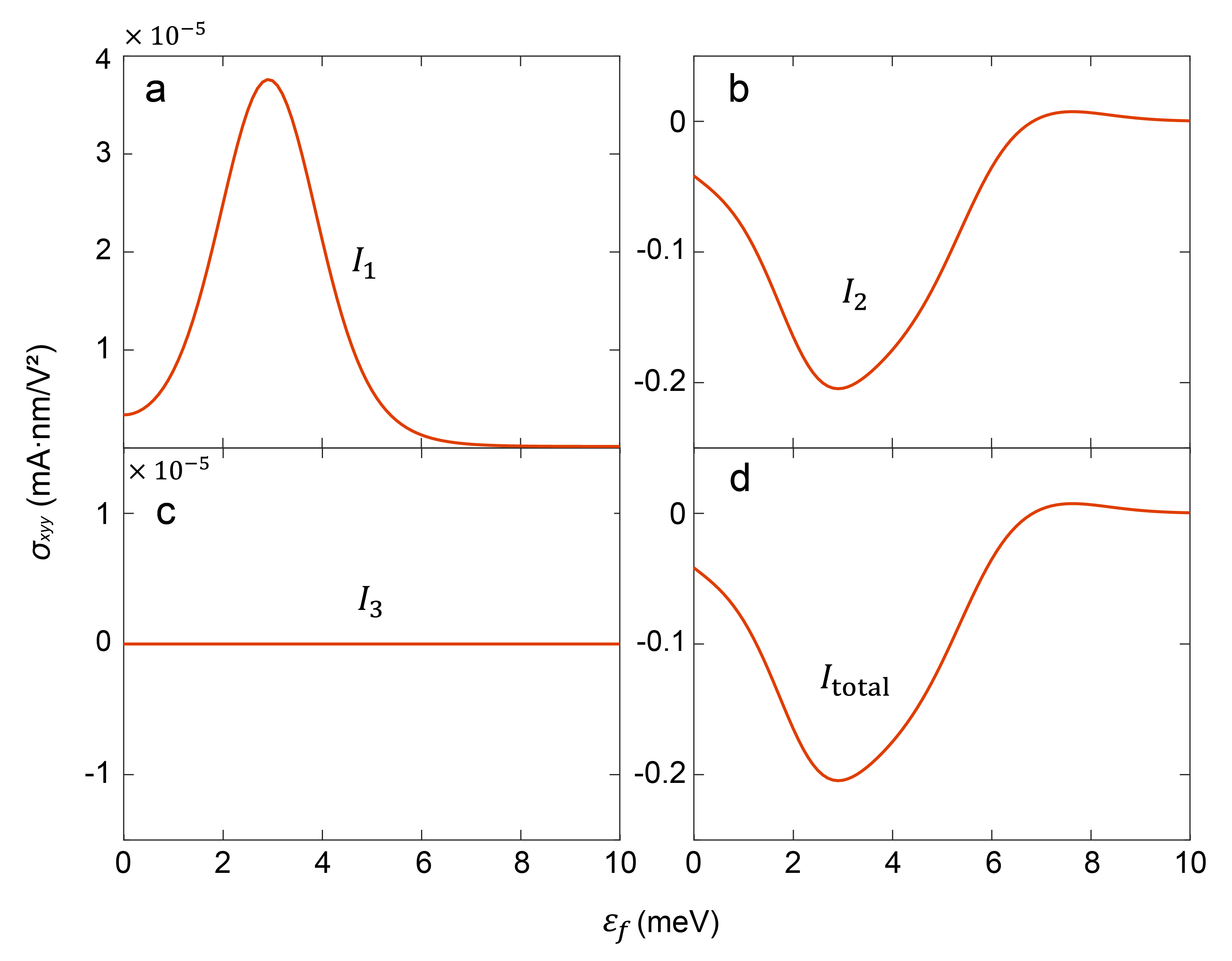}
    \caption{Fermi sphere contribution to $\sigma_{xyy}$ as a function of Fermi energy $\varepsilon_f$ at $V_0 = 2$ meV, $t = 2$meV, and $\Gamma = 0.6$ meV for contributions from (a) $I_1$, (b) $I_2$, (c) $I_3$, and (d) total.}
    \label{sm-fig:fig1}
\end{figure}

Besides the contributions at the Fermi surface in Eq. \eqref{eq:ferimi-surface}, the contributions below the Fermi surface (Fermi sphere or Fermi sea) to the nonlinear conductivity are 
\begin{equation}
    \sigma_{abc} = \frac{e^3}{\hbar} \sum_{n\neq m} \int_k f_n (I_1 + I_2 + I_3)_{nm},
\end{equation}
where
\begin{equation}
\begin{aligned}
    (I_1)_{nm} = \frac{1}{\varepsilon^2_{nm}} &\left[ 2(v^a_{mm} - v^a_{nn}) \mA^b_{nm} \mA^c_{mn} \right. \\
    &\left. - (v^b_{mm} - v^b_{nn}) (\mA^a_{nm} \mA^c_{mn} + \mA^c_{nm} \mA^a_{mn}) \right], \\ 
    (I_2)_{nm} = \frac{1}{\varepsilon_{nm}} &\left[\mA^b_{nm} \partial_c \mA^a_{mn} + \mA^b_{mn} \partial_c \mA^a_{nm} \right. \\
    &- \mA^a_{nm} \partial_c \mA^b_{mn} - \mA^a_{mn} \partial_c \mA^b_{nm} -  \\ 
    &\left. i(\mA^b_{nn} - \mA^b_{mm}) (\mA^c_{nm} \mA^a_{mn} - \mA^a_{nm} \mA^c_{mn}) \right], \\
    (I_3)_{nm} = \sum_{l\neq m \neq n} &\frac{i}{\varepsilon_{nm}} \left[\mA^b_{nm}(\mA^c_{ml} \mA^a_{ln} - \mA^a_{ml} \mA^c_{ln}) \right. \\
    & \left. + \mA^b_{mn}(\mA^c_{nl} \mA^a_{lm} - \mA^a_{nl} \mA^c_{lm})  \right].
\end{aligned}
\end{equation}
$I_1$, $I_2$, and $I_3$ can be understood as the velocity shift, positional shift, and renormalization of the Berry curvature, respectively \cite{kaplan2023general}. We numerically evaluate the Fermi sphere contribution in Fig. \ref{sm-fig:fig1} and find that the Fermi sphere contribution is negligible compared to the Fermi surface contribution.

\bibliography{ref.bib}

\end{document}